\documentclass{PoS}

\newcommand{\calo}{{\cal O}}

\title{Dispersive representation and shape \\ of $K_{\ell 3}$ form factors}

\ShortTitle{Dispersive representation and shape of $K_{\ell 3}$ form factors}

\author{\speaker{Emilie Passemar}\thanks{Work in collaboration with V.~Bernard, M.~Oertel and J.~Stern~\cite{Bernard:2006gy},~\cite{Bernard:2007}.}\\

        Groupe de Physique Théorique, IPN,
        Université de Paris Sud-XI, F-91406 Orsay, France\footnote{Address after the 1st of July : Institute for theoretical physics,
University of Bern, Sidlerstr. 5, CH-3012 Bern, \\~~~~~~~~~~~~~~~~~~~~~~~~~~~~~~~~~~~~~~~~~~~~~~~~~Switzerland}\\
        E-mail: \email{passemar@ipno.in2p3.fr}}

\abstract{The Callan-Treiman low-energy theorem offers an opportunity to test electroweak couplings of light quarks to the gauge boson W. To that aim, 
we introduce a model-independent and accurate dispersive parametrization of the 
two $K_{\ell 3}$ form factors. We 
then discuss three applications to the analysis of $K_{e3}$ and $K_{\mu3}$ measurements:
the prediction of the ratios $\Gamma \left(K_{\mu3}\right)$/$\Gamma\left(K_{e3}\right)$, the extraction of $|f_+(0)V_{us}|$ 
and finally the possible measurement of $m_u-m_d$ induced isospin breaking asymmetry.}

\FullConference{KAON International Conference\\
		 May 21-25 2007\\
		 Laboratori Nazionali di Frascati dell'INFN, Rome, Italy}

\begin{document}
\section{Introduction}
The hadronic matrix element involved in $K_{\ell 3}$ decays is described in terms of two form factors $f_+$ and $f_-$. 
The vector form factor $f^{K\pi}_+ (t)$ represents
the $P$-wave projection of the crossed channel matrix element
$\langle 0 |\bar{s}\gamma_{\mu}u | K\pi \rangle$, whereas                
the $S$-wave projection is described by the scalar form factor
\begin{equation}
f_S^{K\pi}(t) = f^{K\pi}_+ (t) + \frac{t}{m^2_K - m^2_\pi} f^{K\pi}_-(t)~,
\label{defffactor}
\end{equation}
where $t=(p_{\pi}-p_{K})^2$. In the following, we will consider the normalized form factors
$f_0(t)=f^{K^0\pi^-}_S(t)/f^{K^0\pi^-}_+(0)$,~$f_+(t)=f^{K^0\pi^-}_+(t)/f^{K^0\pi^-}_+(0)$, $f_0(0)=f_+(0)= 1,$
and we will try to describe their shape as precisely as possible in the region of interest, in order to test the Standard Model.
\section{Callan-Treiman theorem: a test of the Standard Model}

The low energy theorem of Callan and Treiman~\cite{Dashen:1969bh} predicts the value of $f_0(t)$ 
at the Callan-Treiman (CT) point, namely $t=\Delta_{K \pi}\equiv m_{K}^2-m_{\pi}^2$, in the 
$\mathrm{SU}(2)\times \mathrm{SU}(2)$ chiral limit ($m_u, m_d \rightarrow  0$). When $m_u, m_d \not = 0$, one writes
\begin{equation}
C\equiv f_0(\Delta_{K\pi})=\frac{F_{K}}{F_{\pi}}\frac{1}{f_{+}^{K^0\pi^-}(0)}+  
\Delta_{CT}~,
\label{C}
\end{equation}
where the CT correction $\Delta_{CT} \sim \calo \left( m_{u,d}/4 \pi F_\pi \right)$. 
This correction has been estimated within Chiral Perturbation Theory (ChPT) at next to
leading order (NLO) in the isospin limit~\cite{Gasser:1984ux} with the result
\begin{equation} 
\Delta^{NLO}_{CT}=-3.5~10^{-3}~.
\label{Delta_CT}
\end{equation}
Note that the CT point $\Delta_{K \pi}$ is situated between the physical end point of $K_{\ell3}$ decays \mbox{$t_0\equiv (m_K-m_\pi)^2$} and the $K\pi$ scattering crossed channel threshold $t_{K\pi}\equiv (m_K+m_\pi)^2$. 
$F_{K,\pi}$ are respectively the kaon and pion 
decay constants parametrizing QCD effects. $\Delta_{CT}$ is not enhanced by 
any chiral logarithm, so its value is rather small. A complete study of  $\mathcal{O}(p^6)$ corrections~\cite{Bijnens:2007km} is not available yet.
Howe-ver, we expect that these effects should not change the value of $\Delta^{NLO}_{CT}$ in  
Eq.~(\ref{Delta_CT}) by one order of magnitude. 
Concerning the study of isospin breaking corrections, contributions from the $\pi^0$-$\eta$ mixing in the final state,
usually responsible for large isospin violations, are absent in the neutral $K_{\ell3}$ mode, $K^0\to \pi^-$. 
In the charged mode instead, these effects can easily reach a few percents. Furthermore, the electromagnetic (EM) effects have still to be fully investigated.

Summarising, Eq.~(\ref{C}) allows to predict with good precision the value of 
$f_0(t)$ at the CT point and test the Standard Model (SM) couplings in a sector where they have not been tested so far. More specifically, 
assuming the SM couplings, we can  test Eq.~(\ref{C}) by measuring independently 
the l.h.s and r.h.s of Eq.~(\ref{C}):
first step, in the SM the QCD parameters on the r.h.s of Eq.~(\ref{C})
can be extracted from experimental information, such as
the ratio $\Gamma_{K^+_{\ell 2(\gamma)}}/\Gamma_{\pi^+_{\ell 2 (\gamma)}}$~\cite{Yao:2006px}, 
$|f_+^{K^0\pi^-}(0) V_{us}|$~\cite{Alexopoulos:2004sw} from $K^0\to \pi^- e^+ \nu_e$ and $V_{ud}$~\cite{Marciano:2005ec}, namely
\begin{equation}
C_{SM} =\left|\frac{F_{K^+} V_{us}}{F_{\pi^+} V_{ud}}\right|
\frac{1}{|f_+^{K^0\pi^-}(0) V_{us}|}|V_{ud}|+ \Delta_{CT} = B_{exp} + \Delta_{CT}~, 
\label{CSM}
\end{equation}
with $B_{exp}=1.2438 \pm 0.0040$~. It can be noticed  
that this prediction is independent of the knowledge of $V_{us}$. 
Besides, since experimentally one measures $|{F_{K^+} V_{us}}/{F_{\pi^+} V_{ud}}|$, 
this will introduce in Eq.~(\ref{C}) some small isospin breaking effects due to the difference between $F_{K^+}$ 
and $F_{K^0}$~\cite{Smith:2007}. 
Finally, the l.h.s of Eq.~(\ref{C}), $f_0(m_K^2-m_\pi^2)$, can be 
determined with some care from the $K_{\mu3}$ decay distribution. 

\begin{figure}[h!]
\hspace*{-0.2cm}
\includegraphics*[scale=0.39]{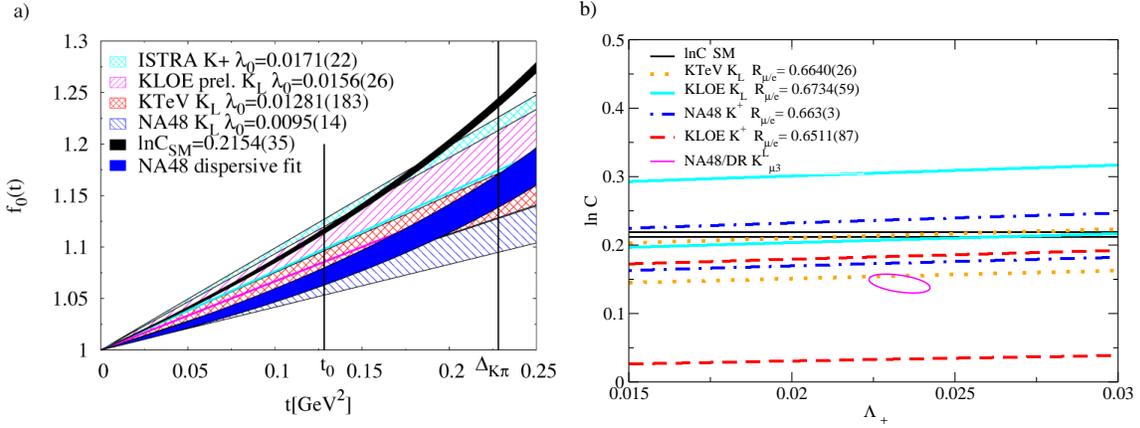}
\hfill
\hspace*{-0.9cm}
\vspace{-0.1cm}
\includegraphics*[scale=0.31]{Br.eps}
\caption{\small{a) Shape of $f_0$ with a linear fit $\left(\lambda_0^{exp}\right)$~\cite{Palutan:2007} compared to the dispersive analysis. b) 1$\sigma$ domains allowed by the branching ratio measurements and by the NA48/DR analysis.}}
\label{Fig1}
\end{figure}

To this end, one has to extrapolate $f_0$ from the physical region $\left( 0<t<t_0\equiv (m_K-m_\pi)^2\right)$ up to $\Delta_{K\pi}$.
Fits to the $K_{\ell 3}$ measured distributions are usually performed using a Taylor expansion of the two form factors 
\begin{equation}
f_{0,+}(t) = 1 + \lambda_{0,+} \left(\frac{t}{m_{\pi}^2}\right) + \frac{1}{2} \lambda_{0,+}' \left(\frac{t}{m_{\pi}^2}\right)^2  + \ldots~.
\label{taylor}
\end{equation}
So far, the results available for the scalar form factor only give the slope $\lambda_0^{exp}$, since the different experiments cannot discriminate $\lambda_0^{exp}$ from 
the curvature 
which are highly correlated. The results disagree among all the experiments as illustrated in Fig.\ref{Fig1}a.
To extrapolate up to $\Delta_{K\pi}$, we cannot neglect the positive curvature. Anyway, even if there was no curvature,  
the SM prediction for the slope implied by the CT theorem 
would be $\lambda_0=0.0208\pm 0.0003$ and that is above all the exis-ting experimental results. Note that the ChPT prediction for $\lambda_0$ is much less precise than the CT theorem since it derives from an expansion in powers of $m_s$. However, the curvature being positive, it is obvious that the measured slope $\lambda_0^{exp}$ can be at most an upper bound for the "true" slope mathematically defined as: $\lambda_0\equiv m_\pi^2 f'(0)$. Under these conditions, $\lambda_0^{exp}$ may depend on the fitted distributions and thus on different weights given to the different areas of the Dalitz plot by various experiments. That could 
explain the apparent discrepancy between the experimental results. 
\section{Dispersive representations of the K$\pi$ scalar and vector form factors} 
\noindent 
\subsection{Scalar form factor}
\begin{figure}[h!]
\begin{minipage}[t]{70mm}
\begin{center}
\includegraphics*[width=6.8cm]{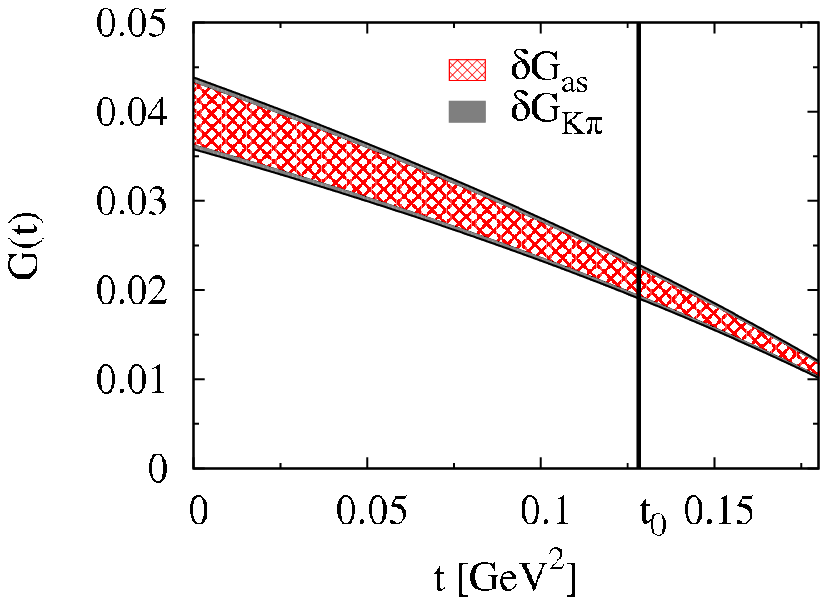}
\vspace{-0.2cm}
\caption{\small{$G(t)$ with the uncertainties $\delta G_{as}$ 
and $\delta G_{K \pi}$ added in quadrature.}}
\label{figureg}
\end{center}
\end{minipage}
\hspace{0.4cm}
\hfill
\begin{minipage}[t]{70mm}
\begin{center}
\hspace{-0.5cm}
\includegraphics*[width=7.2cm]{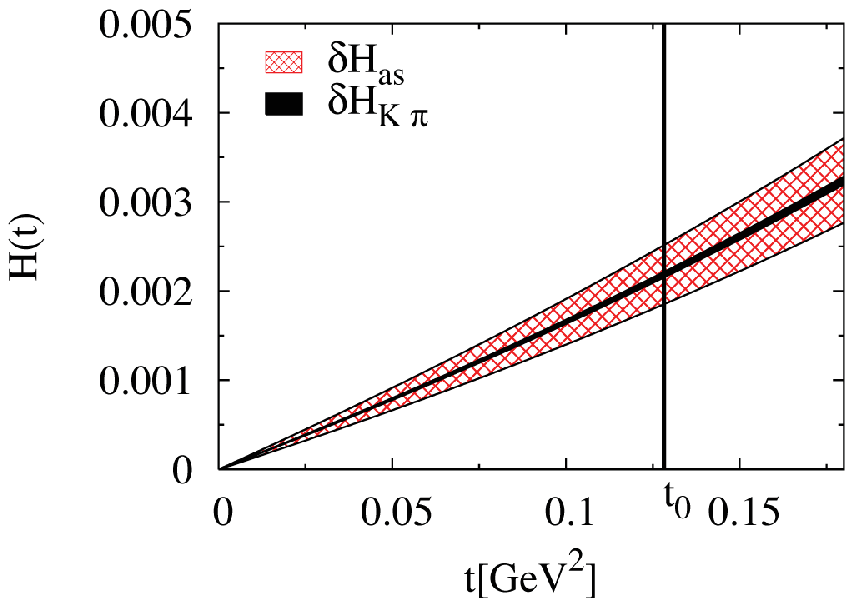}
\vspace{-0.3cm}
\caption{\small{H(t) with the uncertainties $\delta H_{as}$ 
and $\delta H_{K \pi}$ added in quadrature.}}
\label{figureh}
\end{center}
\end{minipage}
\end{figure}

Our aim is to construct a very precise representation of the scalar form factor between 0 and $\Delta_{K\pi}$ 
in order to test the SM prediction at the CT point. The information we have is: the value of $f_0$ at zero,  the K$\pi$ scattering phase in the isospin limit and the fact that $f_0$ has to vanish as $\calo(1/s)$ for large negative s~\cite{Lepage:1979zb}.
In order to determine $C$, one can write a dispersion relation for ln$(f_0(t))$ twice subtracted at 0 and $\Delta_{K\pi}$ 
\begin{equation}
f_0(t)=\exp\Bigl{[}\frac{t}{\Delta_{K\pi}}(\mathrm{ln}C- G(t))\Bigr{]}~,~\mathrm{with}~~ 
G(t)=\frac{\Delta_{K\pi}(\Delta_{K\pi}-t)}{\pi}\nonumber\ \int_{t_{K\pi}}^{\infty}
\frac{ds}{s}
\frac{\phi(s)}
{(s-\Delta_{K\pi})(s-t-i\epsilon)}~,
\label{Dispf}
\end{equation}
where $t_{K\pi}$ is the threshold of $K \pi$ scattering and $\phi(s)$ is the
phase of $f_0(s)$. We decompose the integration range into two parts: the elastic part ($t_{K\pi}<s<\Lambda$) and the (inelastic) asymptotic part ($\Lambda <s<\infty)$ corresponding to the splitting $G(t)=G_{K\pi}(\Lambda,t) + G_{as}(\Lambda,t) \pm \delta G(t)~$, where $\delta G$ represents the total uncertainty. For the elastic part, $G_{K\pi}(\Lambda,t)$, Watson theorem states that in the elastic region and in the isospin limit $\phi(s)$ equals the S-wave, $I=1/2$ K$\pi$ scattering phase $\delta_{K\pi}(s)$. This phase has been extracted (in the isospin limit) from experimental data~\cite{Aston:1987ir} solving the Roy-Steiner equations~\cite{Buettiker:2003pp}. For the asymptotic part, $G_{as}(\Lambda,t)$, we take $\phi(s)=\phi_{as}(s)=\pi \pm \pi$ since $\phi(+\infty)=\pi$ due to the behaviour of $f_0$ at large negative s. Note that we consider a very conservative estimate of the asymptotic uncertainty. 
Thanks to the two subtractions, $G(t)$ in the region of interest is relatively insensitive to the unknown phase at high energy and also to the precise knowledge of $\Lambda$, characterizing the end of the elastic region. In Fig.\ref{figureg} is represented the function $G(t)$ for $\Lambda=2.77~\mathrm{GeV^2}$. 
Let us stress that the uncertainties on $G(t)$ represent at most $10 \%$ of its central value ($G(t)<G(0)=0.0398 \pm 0.0040$) which by itself does not exceed $20 \%$ of the expected value of ln$C$. Hence, one can be confident that the theoretical uncertainty is under control and will 
not be 
the dominant error in the experimental extraction of ln$C$. 
We have sticked here to the isospin limit. 
The potentially dangerous contribution of the cusp at $(m_{K^{+}}+ m_{\pi^{0}})^2$ due to isospin breaking should however not affect our results.  
Besides, a zero of $f_0(t)$ could only influence our dispersive construction if it was located at $t<t_{K\pi}$, close to the physical region. This seems unlikely due to 
$f_0(0)=1$ and the smallness of the slope. For more details on these two last points see Ref.~\cite{Bernard:2007}.\\
\indent In order to determine ln$C$, in principle there exists a sum rule 
\begin{equation}
G(- \infty )=\mathrm{ln}C
\label{Sumrule}
\end{equation}
dictated by the asymptotic behaviour of $f_0$. However, it is not precise enough to allow to determine ln$C$ with the needed accuracy without adding any information on the high energy behaviour of the phase of the form factor. Indeed, the integrand defined in Eq.~(\ref{Dispf}) drops as $1/s^3$ whereas the one in $G(-\infty)$, which has one less subtraction, drops as $1/s^2$.
For instance, if we vary $\Lambda$ between $2.25~\mathrm{GeV^2}$ and $2.77~\mathrm{GeV^2}$, $G(-\infty)$ varies between 0.1335 and 0.3425, whereas $G(t)$ is practically not affected. We will thus instead determine ln$C$ from experiment by fitting the $K_{\mu3}$ decay distribution with the dispersive representation formula of $f_0(t)$, Eq.~(\ref{Dispf}). Once ln$C$ is known, we can estimate the first two coefficients of the Taylor expansion, Eq.~(\ref{taylor}), and their correlation
\begin{equation}
\lambda_0 =\frac{m_{\pi}^2}{\Delta_{K \pi}}(\mathrm {ln C} - G(0) )~~~\mathrm{and}~~~
\lambda'_0 = \lambda^2_0  - 2 \frac{m_{\pi}^4}{\Delta_{K\pi}} G'(0) = \lambda^2_0  + (4.16 \pm 0.50)\times 10^{-4}~. 
\label{slopecurvature}
\end{equation}
\subsection{Vector form factor} 
One can use the same dispersive construction as for the scalar form factor. Writing a twice subtracted representation of the form factor but this time at 0 ($f_+(0)=1$ and $f'_+(0)=\Lambda_+/m^2_\pi$) leads to
\begin{equation}
f_+(t)=\exp\Bigl{[}\frac{t}{m_\pi^2}\left(\Lambda_+ + H(t)\right)\Bigr{]}~,~\mathrm{where}~~ 
H(t)=\frac{m_\pi^2t}{\pi} \int_{t_{K\pi}}^{\infty}
\frac{ds}{s^2}
\frac{\varphi (s)}
{(s-t-i\epsilon)}~.
\label{Dispfp}
\end{equation}
Here, in the elastic region, the phase of the vector form factor $\varphi(s)$ equals the $I=1/2$, P-wave scattering phase. The procedure using Roy-Steiner equations suffers in this case from a lack of relevant experimental inputs. Hence, we have constructed a partial wave amplitude using a Breit-Wigner parametrization around the $K^*(892)$ pole (using the PDG value as input) which is analytic, unitarized and has the correct threshold behavior following 
Ref.~\cite{Gounaris:1968mw}. The resulting function H is decomposed into two parts as previously with $\Lambda \sim (1.4~\mathrm{GeV})^2$, 
see Fig.\ref{figureh}. Note that in this channel the $K^*(892)$ pole dominates and the vector form factor is very well described by the pole parametrization, 
which is in 
full agreement with our dispersive construction.  For other works on the subject, see Ref.~\cite{Hill} and references therein.
\subsection{First dispersive analysis}
NA48 has realized the first dispersive analysis (referred as NA48/DR in the following) of their $K_{\mu3}^L$ Dalitz distribution leading to the result~\cite{Lai:2007dx}:
$\mathrm{ln}C^{exp}=0.1438(140)$ and $\Lambda_+^{exp}=0.0233(9).$ The sum rule $H(-\infty)=-\Lambda_+$ is well fulfilled. However, in the case of ln$C$, ln$C^{exp}$ is close to the lower bound of $G(-\infty)$.
Even if the sum rules are not very stringent since they 
are sensitive to the behaviour of the phase at high energy, the sum rule, Eq.~(\ref{Sumrule}), indicates that 
 the phase of $f_0(t)$ should drop after $\Lambda$ before reaching its asymptotic value $\pi$. Note that a similar phenomenon exists in the case of $\pi \pi$ scattering~\cite{Ananthanarayan:2004xy}. The result for $f_0(t)$ is shown in Fig.\ref{Fig1}a for the SM prediction, Eqs.~(\ref{CSM}) and~(\ref{Delta_CT}), and for the NA48/DR result. 
The NA48/DR analysis indicates a 5$\sigma$ deviation with the SM prediction that could be interpreted as a signal
of physics beyond the SM as for instance a first evidence of a direct coupling of W to right-handed quarks~\cite{Bernard:2006gy, Stern2007}. However, the apparent disagreement between the different experimental measurements has to be clarified before drawing a firm conclusion.\\ 
\indent According to Eq.~(\ref{slopecurvature}), one can infer in the case of the SM value, Eq.~(\ref{CSM}),
$\lambda^{~SM}_0 = 0.01523 \pm 0.00046 + 0.069 \Delta_{CT}$\footnote{This is in complete agreement with the prediction of~\cite{Jamin:2004re}.},$~~\lambda'^{~SM}_0=(6.48 \pm 0.52 + 47.1 \Delta_{CT})~10^{-4}$ and in the case of the NA48/DR result :
$\lambda^{~NA48}_0 = 0.0089 \pm 0.0012,~~\lambda'^{~NA48}_0=(4.95 \pm 0.55)~10^{-4}$~.
\section{Applications}
The knowledge of $f_0(t)$ and $f_+(t)$ allows to determine the phase space integrals $I^\ell_{K^{+/0}}$ 
entering the well-known master formula of the $K_{\ell3}$ partial widths, intensively discussed during the confe-rence (see Talks in the $V_{us}$ session),
\begin{equation}
\Gamma_{K^{+/0}_{\ell 3}}=\mathcal{N}_{K^{+/0}}~S_{EW}~(1+ 2\Delta^{EM}_{K^{+/0}\ell})~|f_+^{K^{+/0}}(0)V_{us}|^2~I^\ell_{K^{+/0}}~,~\mathcal{N}_{K^{+/0}}=C_{K^{+/0}}^2~G_F^2~m^5_{K^{+/0}}/(192 \pi^3).
\label{DecayW}
\end{equation}
Here, we do not consider the isospin breaking (IB) correction, $\Delta_{SU(2)}\equiv f_+^{K^+\pi^0}(0)/f_+^{K^0\pi^+}(0)-1$, mostly induced by $m_u-m_d$ effects, 
as a known theoretical input since we would like to determine it experimentally.
Assuming $\mu/e$ universality, separately tested in the ratio $\Gamma(K_{e2})/\Gamma(K_{\mu2})$~\cite{Wanke:2007}, the knowledge of $I^\ell_{K^{+/0}}$ can be converted into a prediction for $R^{+/0}_{\mu/e}\equiv \Gamma_{K^{+/0}\mu3}/\Gamma_{K^{+/0}e3}$.
Using the last update of the EM corrections~\cite{Cirigliano:2001mk}, we obtain the results in Tab.1(left).
\begin{table}
\hspace{-0.1cm}
\hspace{-0.3cm}
\begin{tabular}{c|c|c}
\hline \hline
$I_{K^{+/0}}^\ell$ & $R^0_{\mu/e}$ & $R^+_{\mu/e}$\\
\hline
$\mathrm{SM}$ & 0.6677(28) &  0.6640(28)  \\
$\mathrm{NA48/DR}$ & 0.6589(33) & 0.6552(33) \\
\hline \hline 
\end{tabular}
\hfill
%
\begin{tabular}{c|c|c|c}
\hline \hline $I_{K^{+/0}}^\mu$ & $\mid f_+(0) V_{us} \mid_{\mu}^{K^0}$& $\mid f_+(0) V_{us} \mid_{\mu}^{K^+}$ & $\Delta_{SU(2)}^{\mu} $ \\ \hline
SM   & 0.21642(60) &  0.22176(100) & 2.47(54)$\%$\\ 
\hline 
NA48/DR  & 0.21834(69) &  0.22378(107) & 2.49(54)$\%$\\ 
\hline \hline 
\end{tabular}
\caption{\small{Left: Prediction of $R^{+/0}_{\mu /e}$. Right: Extraction of $|f_+(0) V_{us}|$ and $\Delta_{SU(2)}$ for the muonic mode using $I_K^\mu$ calculated with the dispersive approach together with the experimental inputs from~\cite{Palutan:2007} and the preliminary updated EM corrections from~\cite{Cirigliano:2001mk}.}}
\end{table}
For the evaluation of $I_{K^{+/0}}^\ell$, we have used in the case of the SM prediction, ln$C$ from Eq.~(\ref{CSM}) with $\Delta_{CT}$ from Eq.~(\ref{Delta_CT}) and $\Lambda_+=0.0245$, corresponding to the $K^*(892)$ pole, and in the case of the NA48/DR result, ln$C^{exp}$ and $\Lambda_+^{exp}$ including the correlation from this analysis.
These calculations can be compared with the experimental results summarized in the legend of Fig.1b.
$R^{+/0}_{\mu/e}$ depends almost exclusively on $\mathrm{ln}C$, since the dependence on $\Lambda_+$ partially cancels by taking the ratio. This offers the possibility of extracting $\mathrm{ln}C$ quasi 
independently of $\Lambda_+$ contrary to the Dalitz plot analysis.
In Fig.\ref{Fig1}b are plotted, in the plane $(\mathrm{ln}C,~\Lambda_+)$, the 1$\sigma$ domains allowed by the individual $K^0$ and $K^+$ branching ratio 
(BR) measurements. 
Note that one loop ChPT calculations~\cite{Cirigliano:2001mk} show that the form factor shapes (ln$C$, $\Lambda_+$) are, to a good approximation, not affected by IB. 
For comparison, we show on the same plot the independent result of the NA48/DR analysis, which is not incompatible with the BR measurements, though on the lower side of the domain.\\ \indent Using the master formula, Eq.~(\ref{DecayW}), we can extract $|f_+(0)V_{us}|$ from the experimental measurements (see the Flavianet fit~\cite{Palutan:2007}). Waiting for other experiments to perform a similar dispersive analysis, the extraction from the muonic mode, Tab.1(right), only involves the NA48/DR result and hence has still a limited accuracy. 
As it has been emphasized during the conference, see Ref.~\cite{Cirigliano:2007}, the measurements of $|f_+(0)V_{us}|$ from the neutral and charged modes as well as the EM corrections~\cite{Cirigliano:2001mk} have now reached a sufficient degree of accuracy 
to give an opportunity to test the theoretical prediction $\Delta_{SU(2)}=0.0231(22)$~\cite{Leutwyler:1996qg} and the input value of $R=(m_s-\hat{m})/(m_d-m_u)$. 
As it has been shown in Ref.~\cite{Palutan:2007} and from Tab.1(right), using an average of the $K^+$ BR measurements, $\Delta_{SU(2)}$ now seems closer to the theoretical prediction.\\
\\
\indent In conclusion, we have introduced a physically motivated model-independent dispersive re-presentation of the $K_{\ell 3}$ form factors which only involves one free parameter. This will improve the accuracy of the $K_{\mu 3}$ decay analysis and can help to partially remove the apparent discrepancy between the different experimental results (see also the discussion in Ref.~\cite{Franzini}). It will then allow to test the SM and to extract $|f_+(0)V_{us}|$ with a better precision.
\\
\\
{\bf Acknowledgements}: I would like to thank the organisers for this very interesting conference leading to very illuminating discussions as well as for their financial support. 
I am indebted to H.~Neufeld for giving me his preliminary determination of the EM corrections. This work has been supported in part by the EU contract MRTN-CT-2006-035482 (Flavianet).

\end{document}